\documentclass{aastex62}

\usepackage{graphicx}
\usepackage{graphics}
\usepackage{amssymb}
\usepackage{bm}
\usepackage{times}

\usepackage{amsmath}
\usepackage{dcolumn}
\usepackage{tabularx}
\usepackage{xcolor}

\newcommand{\kms}{km~s$^{-1}$}

\newcommand{\ms}{\ensuremath{\rm M_{\odot}}}
\newcommand{\ha}{\ensuremath{{\rm H}\alpha}}

\newcommand{\arcs}{\ensuremath{^{\prime\prime}}}

\shorttitle{SLSN PTF10tpz in dense molecular gas}
\shortauthors{Arabsalmani et al.}
\begin{document}
\title{A superluminous supernova in high surface density  molecular gas within the  bar of a metal-rich galaxy   }

\correspondingauthor{Maryam Arabsalmani}
\email{maryam.arabsalmani@unimelb.edu.au}

\author{M. Arabsalmani}, 
\altaffiliation{}
\affiliation{School of Physics, University of Melbourne, Victoria 3001, Australia}
\affiliation{ARC Centre of Excellence for All Sky Astrophysics in 3 Dimensions (ASTRO 3D), Australia}
\affiliation{IRFU, CEA, Universit\'e Paris-Saclay, F-91191 Gif-sur-Yvette, France}

\author{S. Roychowdhury}
\affiliation{Centre for Astrophysics and Supercomputing, Swinburne University of Technology, Hawthorn, VIC 3122, Australia}
\affiliation{Institut d'Astrophysique Spatiale, CNRS, Université Paris-Sud, Université Paris-Saclay, Bât. 121, 91405 Orsay Cedex, France}

\author{F. Renaud}
\affiliation{Department of Astronomy and Theoretical Physics, Lund Observatory, Box 43, 221 00 Lund, Sweden}

\author{D. Cormier}
\affiliation{IRFU, CEA, Universit\'e Paris-Saclay, F-91191 Gif-sur-Yvette, France}

\author{E. Le Floc'h}
\affiliation{IRFU, CEA, Universit\'e Paris-Saclay, F-91191 Gif-sur-Yvette, France}

\author{E. Emsellem}
\affiliation{European Southern Observatory, Karl-Schwarzschildstrasse 2, 85748 Garching Bei Muenchen, Germany}

\author{D. A. Perley}
\affiliation{Astrophysics Research Institute, Liverpool John Moores University, 146 Brownlow Hill, Liverpool L3 5RF, UK}

\author{M. A. Zwaan}
\affiliation{European Southern Observatory, Karl-Schwarzschildstrasse 2, 85748 Garching Bei Muenchen, Germany}

\author{F. Bournaud}
\affiliation{IRFU, CEA, Universit\'e Paris-Saclay, F-91191 Gif-sur-Yvette, France}

\author{V. Arumugam}
\affiliation{European Southern Observatory, Karl-Schwarzschildstrasse 2, 85748 Garching Bei Muenchen, Germany}

\author{P. M\o ller}
\affiliation{European Southern Observatory, Karl-Schwarzschildstrasse 2, 85748 Garching Bei Muenchen, Germany}


\begin{abstract}
We report the  Atacama Large Millimeter/submillimeter Array (ALMA) observations of the metal rich host galaxy of superluminous supernova (SLSN) PTF10tpz, a barred spiral galaxy at $z=0.03994$. We find the  CO(1-0) emission to be confined within the bar of the galaxy. The distribution and kinematics of  molecular gas in the host galaxy  resemble gas flows along two lanes running  from the tips  of the bar towards the galaxy center. These  gas lanes  end in a gaseous structure in the inner region of the galaxy, likely  associated with an  inner Lindblad resonance. The  interaction between the large-scale gas flows in the bar and the gas in the inner region  plausibly leads   to the formation of massive molecular clouds and consequently  massive clusters. This in turn can result in formation of massive stars, and thus the likely progenitor of the SLSN  in a young, massive cluster.  This picture is consistent with SLSN PTF10tpz being located near the inner structure. We  find the molecular gas in the vicinity  of the SLSN to have high surface densities, comparable with those in  interacting  galaxies or  starburst regions in nearby galaxies.  This  lends support to  high densities being  favorable conditions for formation of SLSNe progenitors. 

\end{abstract}




\section{Introduction} 
\label{sec:int}

Superluminous Supernovae (SLSNe) are rare and extremely bright transients \citep[][]{Smith07-2007ApJ...666.1116S, Quimby13-2013MNRAS.431..912Q}. Their  luminous emission remains brighter than the peak magnitude of most supernovae (SNe) for hundreds of days. Their peak visual absolute magnitudes are 10-100 times brighter  than those of typical core-collapse supernovae \citep[e.g.,][]{Quimby11-2011Natur.474..487Q}, making it possible  to detect them  at $z>2$ \citep[][]{Cooke12-2012Natur.491..228C, Curtin19-2019ApJS..241...17C}.    Up to date, the  underlying explosion mechanism behind  these recently identified  energetic events  remains speculative. However, their very large energy releases  suggest that they   arise from the  stars  at the very high mass end of the stellar initial mass function (IMF) with $\rm M \gtrsim 100\,\, M_{\odot}$ \citep[][]{Gal-yam09-2009Natur.458..865G}. 
Studying the close environments of SLSNe  can provide valuable insights  into understanding the physical conditions in which their progenitors form, and thus put  constraints on progenitor models for their formation.

Similarly to the conventional classification of SNe according to their spectroscopic properties,  SLSNe are grouped into two  main subclasses: hydrogen-poor events (SLSN-I), and events with spectroscopic signatures of hydrogen \citep[SLSN-II; see][]{Gal-Yam12-2012Sci...337..927G}. 
Both groups appear  to  trace the ultraviolet  light of their host galaxies which  supports the  association between their progenitors and massive stars \citep{Neill11-2011ApJ...727...15N, Lunnan15-2015ApJ...804...90L}. 
Their  hosts  are  found to be mainly  irregular, compact,  dwarf galaxies, with high star formation rate (SFR) surface densities, and high specific star formation rates    \citep[sSFR,][]{Lunnan13-2013ApJ...771...97L, Lunna14-2014ApJ...787..138L, Lunnan15-2015ApJ...804...90L, Perley16-2016ApJ...830...13P, Angus16-2016MNRAS.458...84A}. Although,  the host galaxies of type II SLSNe  seem to cover  a  larger range in galaxy masses and metallicities compared to  type I SLSN hosts \citep{Neill11-2011ApJ...727...15N, Perley16-2016ApJ...830...13P, Angus16-2016MNRAS.458...84A}.

The typical low metallicity of SLSN hosts   could imply a strong  metallicity bias for SLSN progenitors \citep[][]{Lunnan14-2014ApJ...787..138L, Perley16-2016ApJ...830...13P, Chen17-2017MNRAS.470.3566C}, favouring  single-star progenitor models with low mass loss due to stellar winds \citep[][]{Langer07-2007A&A...475L..19L}.   However  the detection  of a fraction of SLSNe (of both types) in quite large, massive and metal-rich host galaxies raises  questions about whether low metallicity is indeed necessary for the formation of a SLSN progenitors \citep[see][]{Lunnan15-2015ApJ...804...90L, Perley16-2016ApJ...830...13P, Nicholl17-2017ApJ...845L...8N}.  
In some of the metal-rich cases,  the location of the SLSN is coincident with the central regions  -- the most metal-rich regions of the  host galaxy   \citep[][]{Gal-Yam12-2012Sci...337..927G}.  
A more likely, but less addressed, factor  is formation of  SLSN progenitor stars in dense environments. Massive stars are more likely to be found in  massive star clusters which form  in dense  giant molecular clouds (GMCs). 
This picture  is consistent with the occurrence of SLSNe in the central regions of  metal-rich galaxies as gas flows, driven by galaxy  interactions or bars, can create  dense concentrations of molecular gas in the central regions of galaxies \citep[][]{Sheth05-2005ApJ...632..217S, Gallagher18-2018ApJ...858...90G}. 
It is also compatible  with  progenitor models in which    massive  stars in  dense and young  star cluster merge into a star with sufficient mass for  forming a SLSN  \citep[][]{Portegies07-2007Natur.450..388P, Pan12-2012MNRAS.423.2203P, Heuvel13-2013ApJ...779..114V}.

SLSN PTF10tpz (type II SLSN) at $z=0.03994$ is one of the  SLSNe which occurred close to the nucleus of a massive and  metal-rich  galaxy. 
In this letter we  study the properties of   molecular gas, the fuel for star formation, in the close environment  of this SLSN, and investigate whether its progenitor is formed in a dense environment.  
We present deep  carbon monoxide (CO) emission line  observations of the host galaxy of SLSN PTF10tpz  obtained by ALMA.    The high spatial and spectral resolution of our ALMA observations and the  fairly close distance  of the host galaxy  \citep[183 Mpc using   a flat $\Lambda$CDM with cosmological parameters  from][]{Planck18-2018arXiv180706209P}   allow us to model the dynamics of gas in the galaxy and  to investigate the conditions in which the SLSN progenitor formed. This is the first study of molecular gas in a SLSN host galaxy.


\section{Observations and data analysis}
\label{sec:obs}

We used the Band-3 receivers of the ALMA 12-m Array in C43-6 configuration to map the CO(1-0) emission of the host galaxy  on January 04, 2018 (project code: 2017.1.01568.S; PI: Arabsalmani). 
The observations used  a 2 GHz band, centred on 110.883 GHz, covering the redshifted CO(1-0) emission line. This band was sub-divided into 1920 channels, yielding  a velocity resolution of $\sim2.6$ \kms, and a total velocity coverage  of about 5400 \kms. 
The initial calibration of the data is done by the ALMA support staff, using the ALMA data pipeline in the {\sc{Common Astronomy Software Applications (casa)}} package \citep[][]{McMullin07-2007ASPC..376..127M}. 
We use the {\sc{casa}} package {\sc{tclean}} to produce a spectral cube with  a velocity resolution of $\sim$ 16 \kms. With  Natural weighting, we yield  a synthesised beam size of 0.36\arcs$\times$0.52\arcs, and a root-mean-square  noise of 0.3 mJy per channel in the cube. 

We create  a secondary data cube from our spectral data cube by applying Hanning smoothing across  blocks of three consecutive velocity channels (using {\sc{casa-specsmooth}} task), followed by spatial smoothing by convolving with a Gaussian kernel of full width at half maximum  equal to 6 pixels (using {\sc{casa-imsmooth}} task). The smoothing ensures that any localized noise peaks are ignored and only emission correlated spatially and in velocity is chosen.
We then mask  out pixels in the original spectral data cube which lie below a threshold flux in the secondary data cube.
The threshold flux used to select pixels is approximately twice the noise in a line-free channel of the original spectral data cube. We  obtain  the total intensity and the intensity-weighted velocity field maps of the CO(1-0) emission line by applying the {\sc{immoment}} tasks in {\sc{casa}} package on the original spectral data cube with masked pixels.


We also use a R-band image of SLSN PTF10tpz host galaxy presented in \citet[][]{Perley16-2016ApJ...830...13P}. This image was obtained on 26-November-2011 by the Low-Resolution
Imaging Spectrometer \citep[LRIS,][]{Oke95-1995PASP..107..375O}   on the Keck I 10 m telescope (Program PI: S. Kulkarni, proposal ID: C219LA). 
We fit elliptical isophotes to the optical image using the {\sc{stsdas}} package in {\sc{iraf}}, and find  the observed axial ratios of outermost isophotes converge to be 0.42. This yields   an inclination angle of 68$^o$ for the optical disk of the galaxy assuming a typical  intrinsic axial ratio of 0.2 for the stellar disk.

We  use  archival data of the observations of PTF10tpz with   the Near-Infrared Camera on Keck II (NIRC-2) in adaptive optics mode on UT 2010-10-29 between 06:01 and 06:56 UT (Program PI: A. Boden, proposal ID: C246N2L) in order to  establish the location of PTF10tpz.   Five dithered exposures were acquired in each of the $J$, $H$, and $K$' filters.  Each exposure consisted of 5$\times$30s coadds in $H$ and $K$' and 8$\times$30s coadds in $J$.  A basic reduction with NIR sky subtraction and co-addition of dithered frames is performed. We measure   the astrometric offset of PTF10tpz relative to the galaxy nucleus to be 0.25\arcs\,$\pm$ 0.01\arcs\, W and 0.16\arcs\,$\pm$ 0.01\arcs\, N, relying on the telescope orientation keywords in the header. 

\begin{figure*}[t!]
\centering
\includegraphics[width= 1.0 \textwidth]{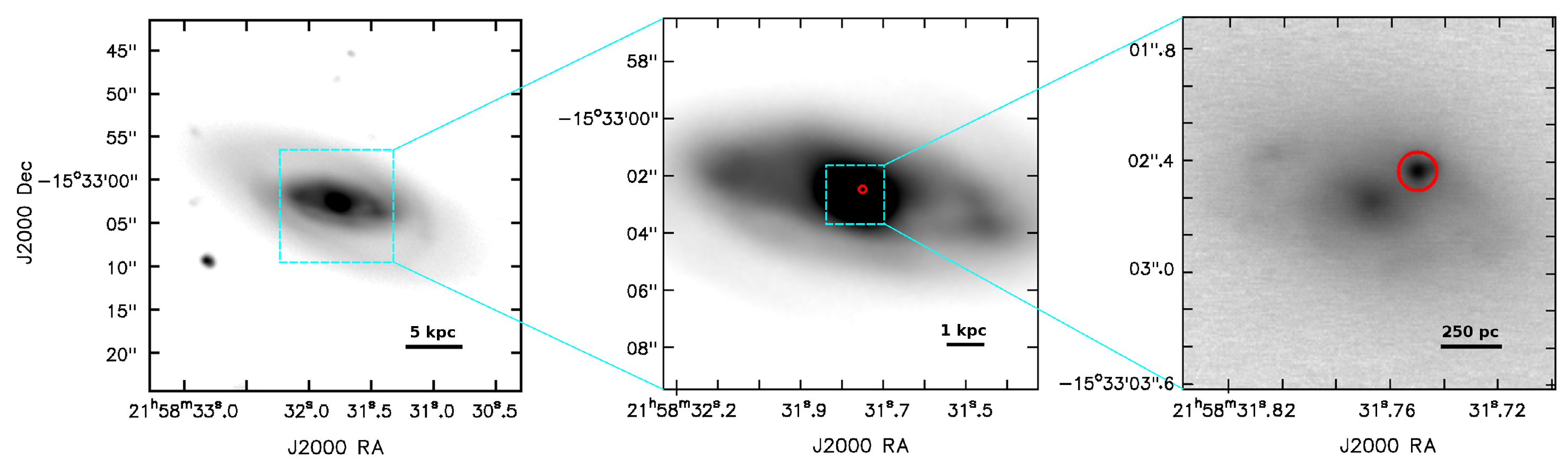}
\caption{{\it{Left and middle panels:}} The R-band image of SLSN PTF10tpz host galaxy obtained by Keck/LRIS showing  the presence of a bar  along with spiral-like  structures in  the galaxy.  The  cyan squares  show  the area  covered in the panel immediately to the right.    \textit{Right panel:} The K-band adaptive optics  image of the  central 1.5 kpc of the host galaxy at the time when SLSN PTF10tpz was bright.  The red circles in middle and right panels  mark the location of the SLSN with a radius indicating  the uncertainties  on the SLSN coordinates.
\label{fig:optical}}
\end{figure*}


\section{Results and Discussion}
\label{sec:res}

\begin{figure*}[t!]
\centering
\includegraphics[width= 1.0 \textwidth]{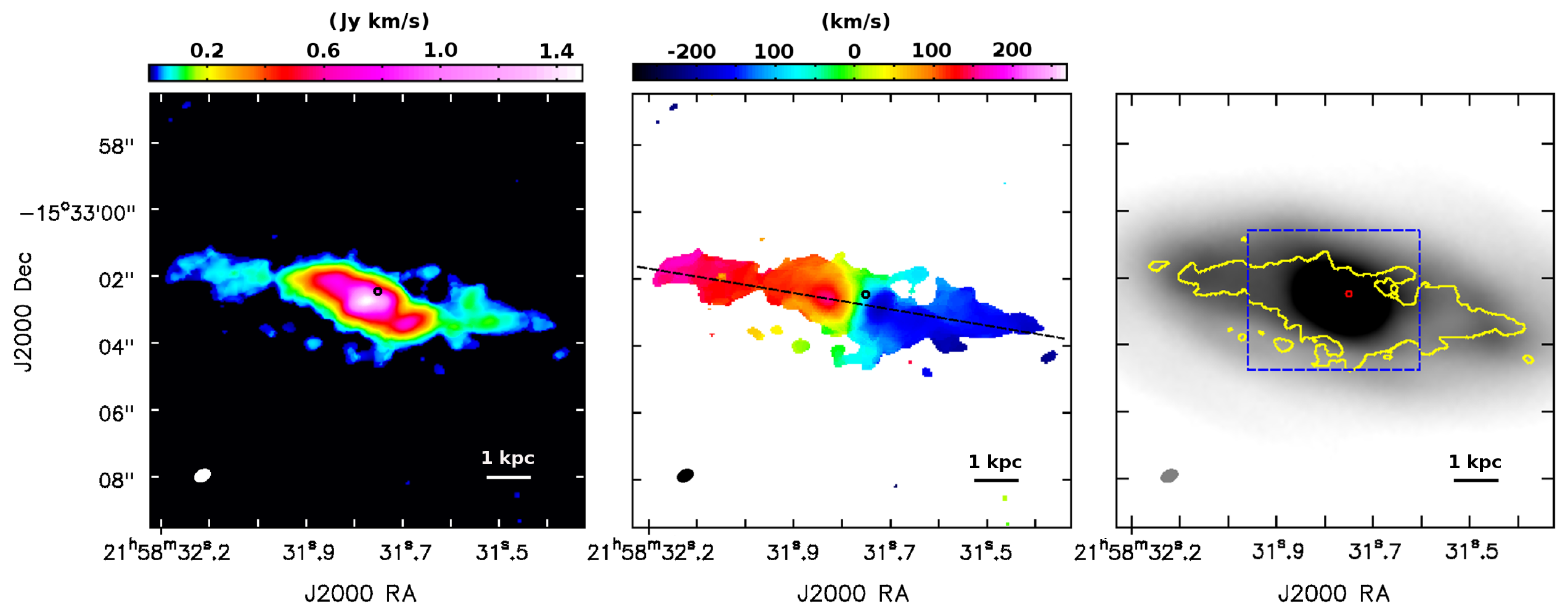}
\caption{  {\it{Left panel:}}  The total intensity  map of  the CO(1-0) emission line in the host galaxy of SLSN PTF10tpz. The colorbar is the flux density of CO(1-0) emission line. 
\textit{Middle panel:} The  intensity-weighted velocity field of the CO(1-0) emission line   corresponding to the left panel. The colorbar is the velocity with respect to the centre of the CO(1-0) emission (at $z=0.03976$). 
{\it{Right panel:}} The R-band  image of the galaxy bar in the same frame as the other two panels, and also the  middle panel  of Fig. \ref{fig:optical}. The yellow contour marks the extent of the CO(1-0) emission at 5$\sigma$ significance. The  dashed blue  box shows the area covered in the  channel maps presented in Fig. \ref{fig:chan-map}.
In all panels the synthesized beam is marked in the bottom-left corner, and the position of the SLSN is marked with a  circle. The radius of the circle indicates the uncertainties  on the SLSN coordinates.  
\label{fig:mom01}}
\end{figure*}

The Keck/LRIS R-band image of the host galaxy of SLSN PTF10tpz is shown in the left and middle  panels of  Fig. \ref{fig:optical}. A bright stellar bar in the centre of the galaxy, along with a faint spiral-like  structure  suggest  that the  host is a SB type  galaxy (see Table \ref{the-only-table} for the global properties of the host).
The right panel of Fig. \ref{fig:optical} shows the inner region of the galaxy at the time when the SLSN PTF10tpz was bright. 
We find  the CO(1-0) emission to be confined  within the  stellar bar of the host galaxy, and extending over a  projected physical size of $\sim$ 10 kpc. The emission has a $\rm W_{20}$ of 470 \kms\, and is centred at a redshift of $z=0.03976 \pm 0.00007$. 
The maps of the total intensity and the intensity-weighted velocity field of the CO(1-0) emission line  are shown in the left and middle  panels of   Fig. \ref{fig:mom01}, respectively. The right panel of Fig. \ref{fig:mom01} shows the R-band image of the galaxy bar within the same frame as in the left and the middle panels (also, same frame as in the middle panel of Fig. \ref{fig:optical}). 

From the total intensity map  it is clear that the CO emission   has a significant contribution from the inner   regions.   Specifically,  the central 1 kpc of the galaxy appears to  contain  about $25\%$   of the total  flux of the CO(1-0) emission.  
The intensity-weighted velocity field of the CO(1-0) emission line    is reminiscent of  a rotating structure   with a distinct  fast rotation   in the  central 1 kpc  of the galaxy. Notably, the molecular gas in this inner region extends over  $> 500$ \kms\ in velocity space.

We measure a velocity-integrated flux density of $18.39\pm0.32$ Jy\,\kms\, over  a total velocity range of 570 \kms, for the CO(1-0) emission from the galaxy. This implies a   brightness temperature luminosity  \citep[as defined in][]{Obreschkow09-2009ApJ...702.1321O}  of $L^{\rm T}_{\rm CO(1-0)}= (1.45\pm0.03) \times 10^{9}\, \rm K\,km\,s^{-1}\,pc^2$. 
We  assume a  Galactic  CO-to-molecular-gas conversion factor  of $\alpha_{\rm CO} = 4.36 \, \rm M_{\odot}\,(\rm K\,km\,s^{-1}\,pc^2)^{-1}$, based on the super-solar metallicity of the host \citep[][]{Perley16-2016ApJ...830...13P} and obtain    a molecular gas mass of $10^{9.8}\,\rm M_{\odot}$ for the host galaxy.  With the stellar mass of $\rm 10^{10.85}\,M_{\odot}$ \citep[][]{Perley16-2016ApJ...830...13P}, this yields a molecular gas mass--to--stellar mass ratio of  $ \sim 0.1$ for the SLSN host, typical of  nearby star forming galaxies with similar stellar masses \citep[][]{Bothwell14-2014MNRAS.445.2599B}. We note that the conversion factor in the central kpc of nearby galaxies is found to be a factor of 2 lower than the mean value \citep[][]{Sandstrom13-2013ApJ...777....5S}. Since a significant  fraction of the CO emission  is from the central few kpc of the host galaxy, its  molecular gas mass  could be lower  by a factor of 2.    
However, for consistency and to compare our results with  the literature we  use the molecular gas mass estimated by assuming a Galactic conversion factor. This does not affect our conclusions.

\setcounter{table}{0}
\begin{table*}
\centering
\caption{{Global  properties of the host galaxy of SLSN PTF10tpz.}
\label{tab:table1}}
\begin{tabular}{cccccccc}
\hline
 RA        & Dec        & redshift & 12+$\rm log_{10}[O/H]$ & $M_*$             & $(S\Delta v)_{\rm CO(1-0)}$ & $\rm W_{\rm 20, CO(1-0)}$& $M_{\rm mol}$   \\
 (J2000)   &  (J2000)   &          &                        & ($\rm M_{\odot}$) & ($\rm Jy\,km s^{-1}$) & (\kms)&  ($\rm M_{\odot}$)   \\
\hline
21:58:31.77  & -15:33:02.61 & 0.03976$\pm$0.00007 & 9.22  & $10^{10.85}$ & ($18.39\pm 0.32$) & 470 & $10^{9.8}$   \\
\hline
\end{tabular}
\vskip 2 mm
\flushleft  {Columns 1 and 2: the coordinates of the galaxy center obtained from the Keck/LRIS image. Column 3: the redshift of the galaxy measured from the CO(1-0) emission line. Columns 4 and 5: the metallicity and stellar mass from \citet[][]{Perley16-2016ApJ...830...13P}. Columns 6 and 7: the velocity-integrated flux density and the $\rm W_{20}$ of the CO(1-0) emission line. Column 8: the molecular gas mass of the host galaxy.} 
\label{the-only-table}
\end{table*}

\citet[][]{Perley16-2016ApJ...830...13P}  performed spectroscopic observations for the host galaxy using a  1 \arcs\ slit passing through the galaxy centre along its major axis and  measured the fluxes of the  bright emission lines in the central regions of the galaxy (within the 1 \arcs\, slit). They  performed a BPT diagram analysis using the log[O{\sc{iii}}/H$\beta$] vs. log([N{\sc{ii}}]/H$\alpha$] diagram  and found the emission line fluxes  to be consistent with the presence of an active galactic nucleus  (AGN) in the galaxy centre. They reported an average  \ha-based SFR  of $\rm 6.0\,M_{\odot}\,yr^{-1}$ for the central regions of the host and stated this to be an upper limit due to a (subdominant) contribution from the  AGN to the  \ha\, flux. 
However, we  note  that the host galaxy on the  log[O{\sc{iii}}/H$\beta$] vs. log([N{\sc{ii}}]/H$\alpha$] plane lies in a region which is at the border between the AGN and star forming galaxy regimes, and hence it is hard to confirm or rule out the presence of an AGN based on this diagram. On the other hand,  the fluxes of the S[{\sc{ii}}] lines reported by \citet[][]{Perley16-2016ApJ...830...13P}  place  the host galaxy amongst  the star-forming galaxies on the   log[O{\sc{iii}}/H$\beta$] vs. log([S{\sc{ii}}]/H$\alpha$] plane \citep[see][]{Kewlwy06-2006MNRAS.372..961K, Singh13-2013A&A...558A..43S}. 
We therefore use the   $\rm 6.0\,M_{\odot}\,yr^{-1}$ as the reference value for the  SFR in the central regions of the galaxy. 
By  measuring  the  molecular gas mass in the same region we estimate an average    molecular-gas-depletion time  ($M_{\rm mol}/\rm SFR$) of 0.7 Gyr,  similar  to depletion times   in nearby starburst galaxies \citep[see e.g.,][]{Bigiel08-2008AJ....136.2846B}.

We obtain a mean      molecular gas surface-density  of   $\Sigma_{\rm mol} \sim 160\,\rm M_{\odot} pc^{-2}$ for the galaxy,  averaged    over the area within which CO(1-0) emission is detected (as defined based on the total intensity map) and  corrected for the inclination of the host.   This is comparable  with the  mean  molecular gas surface-densities of starburst galaxies in the local Universe \citep[also determined by averaging  within the radius of the central molecular disk as determined from their CO maps; see][]{Kennicutt98-1998ApJ...498..541K}.  In order to investigate the nature of the high mean surface density of the host,  we measure  the molecular gas surface-densities over individual  beams, or equivalently over $\sim$ 350 pc scales. We find these to be as   high as  $\sim 1400\,\rm M_{\odot} pc^{-2}$ in the inner region of the galaxy. In particular, we  measure  a  $\Sigma_{\rm mol}$ of  $\sim 700\,\rm M_{\odot} pc^{-2}$  over a beam centred on the SLSN location. These values   are larger or comparable to the molecular gas surface densities obtained for  the central regions of strongly barred nearby galaxies over smaller scales \citep[e.g., 120 pc scales,][]{Sun18-2018ApJ...860..172S}. 
We use the molecular gas clumping  factors from \citet[][]{Leroy13-2013ApJ...769L..12L} to convert  the surface densities averaged over the scales probed by a single beam  to the  surface densities on the scale of individual clouds ($\sim$ 50 pc).  We predict the GMCs in the inner region  of the host galaxy to have surface densities as high as $\sim 5000-10,000\,\rm M_{\odot} pc^{-2}$. These are comparable with  the  surface densities   of GMCs in the Antennae  galaxy or the nuclear starburst of NGC 253    \citep[][]{Leroy15-2015ApJ...801...25L, Sun18-2018ApJ...860..172S}.  The high  molecular gas surface densities in the vicinity of the SLSN supports the hypotheses in which high densities are favorable conditions for formation of SLSNe progenitors \citep[see also][]{Arabsalmani15-2015MNRAS.454L..51A, Roychowdhury19-2019arXiv190300477R, Arabsalmani19-2019arXiv190300485A}. 

\begin{figure*}[t!]
\centering
\includegraphics[width=1.0 \textwidth]{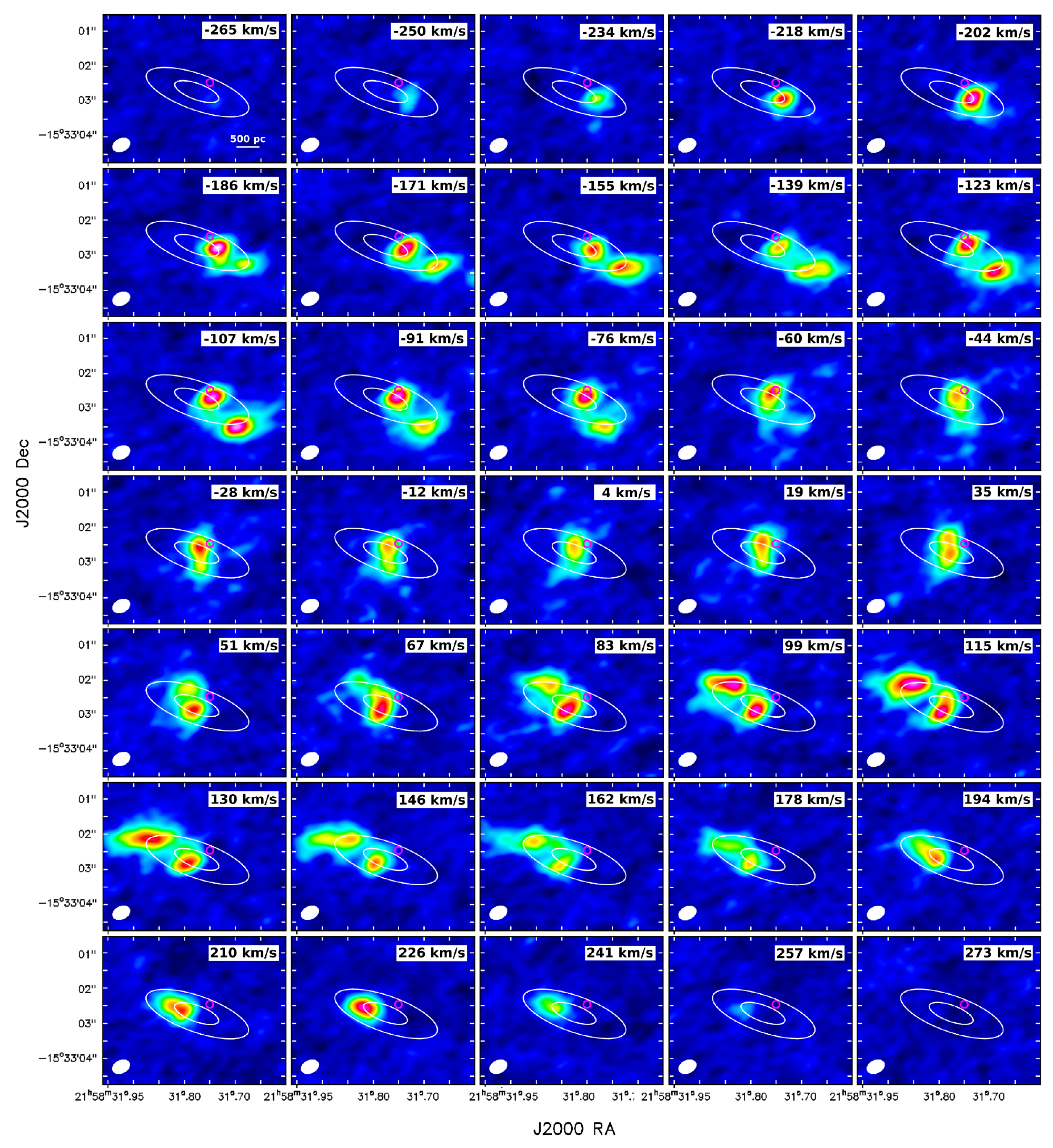}
\caption{The CO emission flux  in successive 16 \kms\, velocity channels. Each map covers the  area marked with the  dashed blue  box in the right panel of Fig. \ref{fig:mom01}. The two white ellipses  are there to help follow the path of the emission through the velocity channels, within the central few kpc of the galaxy. The orientations and  axial ratios of the two ellipses are same as those of the outermost isophotes of the optical disk of the galaxy.   The position of the SLSN is marked with a magenta circle with a radius indicating  the uncertainties  on the SLSN coordinates. 
The beam is shown in the bottom-left corner of each panel. 
\label{fig:chan-map}}
\end{figure*}

\begin{figure*}[t!]
\centering
\includegraphics[width=0.7 \textwidth]{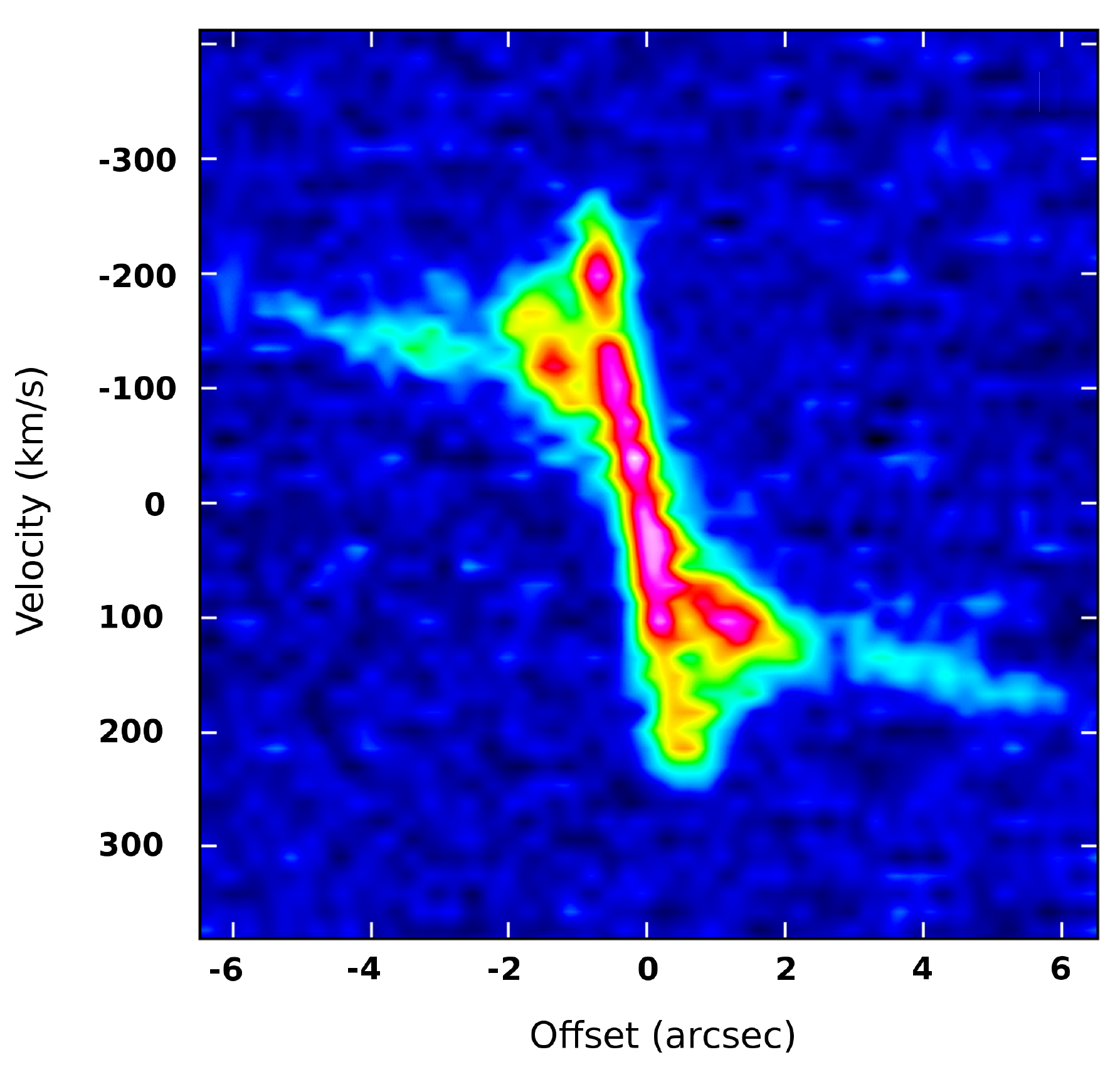}
\caption{The P-V diagram of the CO(1-0) emission where the position varies along the  main rotation axis of the CO emission   (the dashed black line in the middle   panel of Fig. \ref{fig:mom01}).   
\label{fig:pv}}
\end{figure*}

We investigate the kinematics and structure of molecular gas  using channel maps of the CO(1-0) emission at 16 \kms\, velocity resolution presented in  Fig. \ref{fig:chan-map}. These maps cover the inner region of the bar. 
The two white ellipses, following the orientation and flattening of the outermost isophotes of the disk, are there to help follow the path of the emission through the velocity channels. The CO emission is mainly concentrated along the path traced by these two ellipses. 
The emission coincident with the outer ellipse is confined mainly  to the top-left and bottom-right quadrants of the  ellipse. This emission extends beyond the frame  covered in the channel maps in several channels, forming the faint extended wings  which can be seen in the left panel of Fig. \ref{fig:mom01}.  The emission coincident with the inner ellipse on the other hand is mainly restricted to the top-right and bottom-left quadrants of the ellipse,   
and  is present in all channels with velocities between -250 \kms\, and 257 \kms. 
This emission forms a steep profile in the  Position-Velocity (P-V) diagram of the CO(1-0) emission  (Fig. \ref{fig:pv}), making an X-shape pattern when combined with the emission from the outer structure \citep[see][]{Bureau99-1999AJ....118..126B, Merrifield99-1999A&A...345L..47M}.

\begin{figure*}[t!]
\centering
\includegraphics[width=0.7 \textwidth]{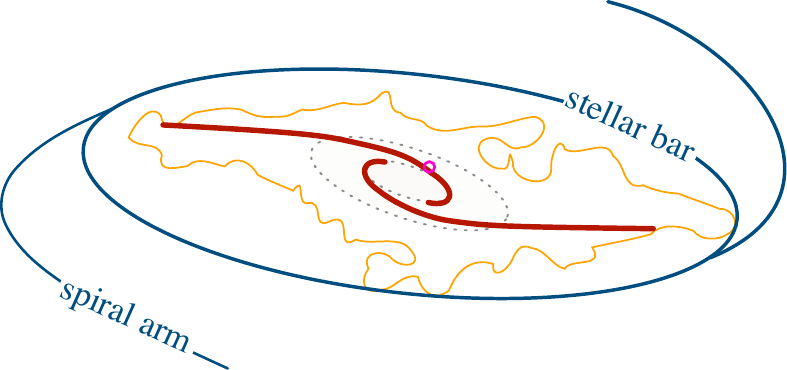}
\caption{Sketch of the central structures in the host galaxy of SLSN PTF10tpz. The  blue ellipse shows the stellar bar of the galaxy. The orange  contour shows the extent of the molecular gas in the bar, as presented in the right panel of Fig. \ref{fig:mom01}. The two dotted gray  ellipses represent the two ellipses in Fig. \ref{fig:chan-map}. The two red spirals   show the gas lanes  along which the molecular gas flows towards the galaxy centre.  The position of the SLSN is marked by a magenta circle.     
\label{fig:cartoon}}
\end{figure*}

The spatial distribution and  kinematics  of the molecular gas  shown in Figures \ref{fig:mom01} and  \ref{fig:chan-map}  
are consistent with  the presence of two gas lanes, running from the tips of the bar towards the galaxy centre. The specific orbital structures imposed by the bar, and the dissipative nature of the gas lead  to two symmetric well-defined gas lanes, often associated with shocks \citep[see e.g.,][]{Maciejewski02-2002MNRAS.329..502M, Emsellem15-2015MNRAS.446.2468E}.  Fig. \ref{fig:cartoon} shows  a simple sketch of such structures  in the bar of the  host galaxy of SLSN PTF10tpz.  
In this picture, the innermost parts of the  two gas lanes   encounter an inner  Lindblad resonance associated with the pattern-speed of the bar, close to the location of the SLSN. The transition of gas from the gas lanes to the inner  resonance  results  in the  formation of high density gas with large velocity dispersions. This in turn leads to the formation of large GMCs  at the intersection of the gas lanes and the inner resonance \citep[similar to what is seen in NGC 1365,][]{Elmegreen09-2009ApJ...703.1297E}. 
The Keck/NIRC2 Kp-band image of the inner region  of the host (the right panel of  Fig. \ref{fig:optical}) suggests the presence of a stellar ring with similar dimensions as those of the inner ellipse in Fig. \ref{fig:chan-map}. 
With a semimajor axis of about 0.1 times the length of the bar, the size of this ring is consistent with the observed median value of 0.11 for the ratio of the nuclear ring size to  the bar size  in the catalog  of \citet[][]{Comeron10-2010MNRAS.402.2462C}.   
This lends support to the association of the molecular gas in the inner region of the galaxy with a resonance  ring. Nuclear rings  are  common in spiral bar galaxies, and are  shown to be the sites of intense starbursts where   massive and compact   star clusters can form \citep[][]{Elmegreen94-1994ApJ...425L..73E, Benedict02-2002AJ....123.1411B, Elmegreen09-2009ApJ...703.1297E, Combes13-2013A&A...558A.124C}. 
Note that deeper  observations with higher spatial resolutions  are required in order to precisely determine the location of the resonance. The  existing  observations do not provide enough information on the pattern speed of the bar, the epicycle frequency, and the mass distribution, to allow a detailed dynamical modeling of the galaxy.

Although  the spatial resolution of data presented here does not allow us to reveal the  exact   nature of the inner structure of molecular gas in the host galaxy, it is likely that the interaction between the large-scale gas flows in the bar and the  gas in the inner region  has led   to cloud-flow or cloud-cloud collisions.  
Observations and simulations  show that cloud-cloud collisions lead to the assembly of giant molecular complexes, and consequently  to the formation of massive star cluster(s) \citep[e.g.,][]{Inoue13-2013ApJ...774L..31I}. This is accompanied by  an increase in the SFR \citep[e.g.,][]{Tan00-2000ApJ...536..173T}, and even a decrease in the depletion time \citep[][]{Renaud15-2015MNRAS.454.3299R}, similar to starbursts, like those typically hosted by interacting galaxies. In addition, it has been proposed that cloud-cloud collisions trigger the formation of massive stars \citep[][]{Motte14-2014A&A...571A..32M, Takahira18-2018PASJ...70S..58T}.

All these, together with the high molecular gas surface densities and short depletion time we report,  suggest  that massive clusters have  formed near the intersection regions of the  gas lanes  and the inner structure that we have identified in the host galaxy of SLSN PTF10tpz. 
Such clusters are  the birth-place of massive stars, potential progenitors of SLSNe, either directly and/or through a runaway process in their core \citep[][]{Portegies07-2007Natur.450..388P, Pan12-2012MNRAS.423.2203P, Heuvel13-2013ApJ...779..114V}. 
These clusters would then continue their motion along the ring (clockwise in Fig. \ref{fig:cartoon}), which is compatible with the location of the SLSN (see Fig. \ref{fig:chan-map} and Fig. \ref{fig:cartoon}). The orbital velocity along the ring together with the separation between the formation and explosion sites of the SLSN progenitor can provide constraints on the lifetime and thus the mass of the progenitor star. 

In order to have a rough estimate for the mass of the progenitor star we assume the orbital velocity along the ring to be same as the radial velocity at the apocenters of the inner ellipse in Fig. \ref{fig:chan-map}, i.e. about 250 \kms.  If the progenitor star was indeed formed at the  intersection  of the  gas lanes  and the inner structure, it  would have  an age of about 3 Myr when it reached the explosion site. This would predict a stellar mass of $\sim$ 100 \ms\, for the SLSN progenitor star \citep[see][]{Leitherer99-1999ApJS..123....3L, Crowther10-2010MNRAS.408..731C}. 
It is however likely that a delay, of the order of a few times the free-fall time,   exists between the assembly of the dense cloud at this intersection and the formation of the star itself. 
For a surface density of a few thousands $\rm M_{\odot} pc^{-2}$ on the scale of individual molecular clouds, we estimate  a volume density of $\sim 50\, \rm M_{\odot} pc^{-3}$ and thus a free-fall time of $\sim 1$ Myr. Considering this delay, the estimated 100 \ms\, is in fact a lower limit for the mass of the  SLSN progenitor star.  
The uncertainty on the  progenitor mass dominantly comes from  the mass--lifetime relation of  very massive stars. 
The theoretical models for this relation  suffer from substantial uncertainties, in particular  due to dependency on parameters such as  the spin and the metallicity of the star and also the  internal mixing \citep[e.g.,][]{Decressin07-2007A&A...464.1029D}. However, according to our current knowledge and state of the art models, timescales of the order of 1-2 Myr correspond to the most massive stars known \citep[see e.g.,][]{Kohler15-2015A&A...573A..71K}. The very short timescale that we estimate between the onset of formation and the explosion of the progenitor star ($\lesssim$ 1 Myr) therefore indicates  that the progenitor star of SLSN PTTF10tpz is amongst the most massive stars.

\section{Summary and Conclusions}
\label{sec:sum}
We study the spatial distribution and kinematics of molecular gas in the metal-rich and barred  host galaxy of SLSN PTF10tpz   though  CO(1-0) emission line observations with ALMA.  
We detect  the CO(1-0) emission line within the bar of the galaxy. The  molecular gas in the inner region of the bar, where the SLSN is located,  have  surface densities as high as $\sim 1400\,\rm M_{\odot} pc^{-2}$ at $\sim 350$ pc scales. This predicts surface densities of $\sim 5000-10,000\,\rm M_{\odot} pc^{-2}$ for individual GMCs, comparable with those    of GMCs in the Antennae  galaxy or the  starburst regions in the nearby Universe. We find the  distribution and kinematics of gas  to be  consistent with two gas lanes running from the tips of the bar towards the galaxy center. These  lanes encounter a gaseous structure in the central $\sim$ 1 kpc of the galaxy. 
The position of SLSN PTF10tpz is very close to the the intersection regions of the gas lanes and the inner structure.  This structure is plausibly associated with an inner Lindblad resonance ring, also suggested by the Keck/NIRC2 image of the inner region of the host.   It is likely that the interaction between the large-scale gas flows in the bar and the gas at the inner resonance leads  to  the assembly of giant molecular complexes, and consequently to the formation of massive star clusters. 
This is  supported   by the short depletion time and high surface densities of molecular gas that we measure in the central regions of the galaxy. 
Our findings  therefore suggest    in-situ formation of massive star clusters (in one of which the SLSN progenitor was formed)  due to the internal  dynamics of the host galaxy and    without any external contribution such as interaction or merger.  This   demonstrates   the  important role that the kpc-scale (hydro-)dynamics of galaxies could play  in the formation of massive stars and hence massive star explosions.

\acknowledgements 
We would like to thank   Nissim Kanekar for valuable contributions to this project.  We  thank Robert Quimby, Andy Boden, and Gaspard Duchesne for triggering and acquiring the Keck/NIRC2 observations. Many thanks to  Ariane Lancon and Fabrice Martins for valuable insights on the lifetime of massive stars. We also thank Peter Erwin for helpful discussions.  
M.A. acknowledges support from  the Australian Research Council Centre of Excellence for All Sky Astrophysics in 3 Dimensions (ASTRO 3D)  through project number CE170100013. 
F.R. acknowledges support from the Knut and Alice Wallenberg Foundation. 
D.C. is supported by the European Union's Horizon 2020 research and innovation programme under the Marie Sk\l{}odowska-Curie grant agreement No 702622. 
ALMA is a partnership of ESO (representing its member states), NSF (USA) and 
NINS (Japan), together with NRC (Canada), NSC and ASIAA (Taiwan), and KASI (Republic of Korea), 
in cooperation with the Republic of Chile. The Joint ALMA Observatory is operated by ESO, AUI/NRAO 
and NAOJ.


\end{document}